\documentclass[reprint,superscriptaddress,preprintnumbers,bibnotes,amsmath,amssymb,aps,prl]{revtex4-1}

\usepackage{graphicx}
\usepackage{dcolumn}
\usepackage{bm} 
\usepackage{color}

\begin{document}


\title{Low-energy spin excitations in optimally doped CaFe$_{0.88}$Co$_{0.12}$AsF superconductor studied with inelastic neutron scattering}

\author{Mingwei~Ma}
\email[]{mw\_ma@iphy.ac.cn}
\affiliation{Beijing National Laboratory for Condensed Matter Physics, Institute of Physics, Chinese Academy of Sciences, Beijing 100190, China}

\affiliation{Songshan Lake Materials Laboratory, Dongguan, Guangdong 523808, China}

\author{Philippe~Bourges}
\email[]{philippe.bourges@cea.fr.}
\affiliation{Universit\'{e} Paris-Saclay, CNRS, CEA, Laboratoire L\'{e}on Brillouin, 91191, Gif-sur-Yvette, France}

\author{Yvan~Sidis}
\affiliation{Universit\'{e} Paris-Saclay, CNRS, CEA, Laboratoire L\'{e}on Brillouin, 91191, Gif-sur-Yvette, France}
\author{Alexandre ~Ivanov}
\affiliation{Intitut Laue Langevin, 71 avenue des Martyrs, CS 20156, 38042 Grenoble Cedex, France}

\author{Genfu~Chen}
\affiliation{Beijing National Laboratory for Condensed Matter Physics, Institute of Physics, Chinese Academy of Sciences, Beijing 100190, China}

\author{Zhian~Ren}
\affiliation{Beijing National Laboratory for Condensed Matter Physics, Institute of Physics, Chinese Academy of Sciences, Beijing 100190, China}
\author{Yuan~Li}
\affiliation{International Center for Quantum Materials, School of Physics, Peking University, Beijing 100871, China}

\begin{abstract}
There are few inelastic neutron scattering (INS) reports on the superconducting single crystals of FeAs-1111 system, even though it was first discovered in 2008, due to the extreme difficulty in large single crystal growth. In this paper, we have studied the low-energy spin excitations in the optimally electron-doped  CaFe$_{0.88}$Co$_{0.12}$AsF single crystals with  $T_\mathrm{c}$  = 21 K by INS. The resonance energy of the superconducting spin resonant mode with $E_\mathrm{r}$ = 12 meV amounts to 6.6 $k_\mathrm{B}$$T_\mathrm{c}$, which constitutes the largest $E_\mathrm{r}$/$k_\mathrm{B}$$T_\mathrm{c}$ ratio among iron-based superconductors reported to date. The large ratio implies a strong coupling between conduction electrons and magnetic excitations in CaFe$_{0.88}$Co$_{0.12}$AsF. \color{black}The resonance possesses a magnonlike upward dispersion along transverse direction due to the anisotropy of spin-spin correlation length within $ab$ plane in the normal-state, which points to a spin fluctuation mediated sign-reversed ${s}\mathbf\pm$ wave pairing in CaFe$_{0.88}$Co$_{0.12}$AsF.  

\end{abstract}

\pacs{74.70.Xa, 
78.70.Nx, 
74.20.Rp, 
74.25.Ha  
}

\maketitle
Neutron spin resonance is commonly observed in the superconducting (SC) state of  copper-oxide \cite{RossatMignod,PhysRevLett.70.3490}, heavy-fermion \cite{SatoNature2001,PhysRevLett.100.087001}, iron-pnictide and iron-chalcogenide superconductors \cite{ChristiansonNature2008,LiuNatureMater2010}. The neutron spin resonance is a collective magnetic excitation occurring below the SC critical temperature $T_\mathrm{c}$ with temperature dependence of its intensity similar to the SC order parameter \cite{Eschrig2006}. The energy of the resonance  $E_\mathrm{r}$ at antiferromagnetic wave vector $\mathbf{Q}_\mathrm{AF}$ is found to be proportional to $T_\mathrm{c}$ for a variety of unconventional superconductors \cite{YuNaturePhys2009,PhysRevB.83.214520}. The most common interpretation of the resonance is that it is a spin exciton, which arises from particle-hole type excitations involving nearly parallel, or nested, sections on the Fermi surfaces that possess opposite signs of the superconducting pairing \cite{Eschrig2006,SIDIS2007745}. In iron-based superconductors the spin resonance is regarded as the hallmark of sign-reversed  ${s}\mathbf\pm$ wave pairings with Bogoliubov quasiparticles in the superconducting state connected by $\mathbf{Q}_\mathrm{AF}$ \cite{PhysRevB.78.020514,PhysRevB.78.144514,PhysRevB.78.140509,Wang2011Sci,Hirschfeld_2011,PhysRevLett.101.206404,MAZIN2009614,PhysRevB.78.134512,Richard_2011} and reveals an upward magnonlike dispersion \cite{PhysRevB.98.060502,PhysRevLett.110.177002}. Moreover, whether the corresponding spin excitations above $T_\mathrm{c}$ have sufficient spectral weight can reveal their qualification to be regarded as pairing bosons.

\begin{figure}
\includegraphics[width=3.5in]{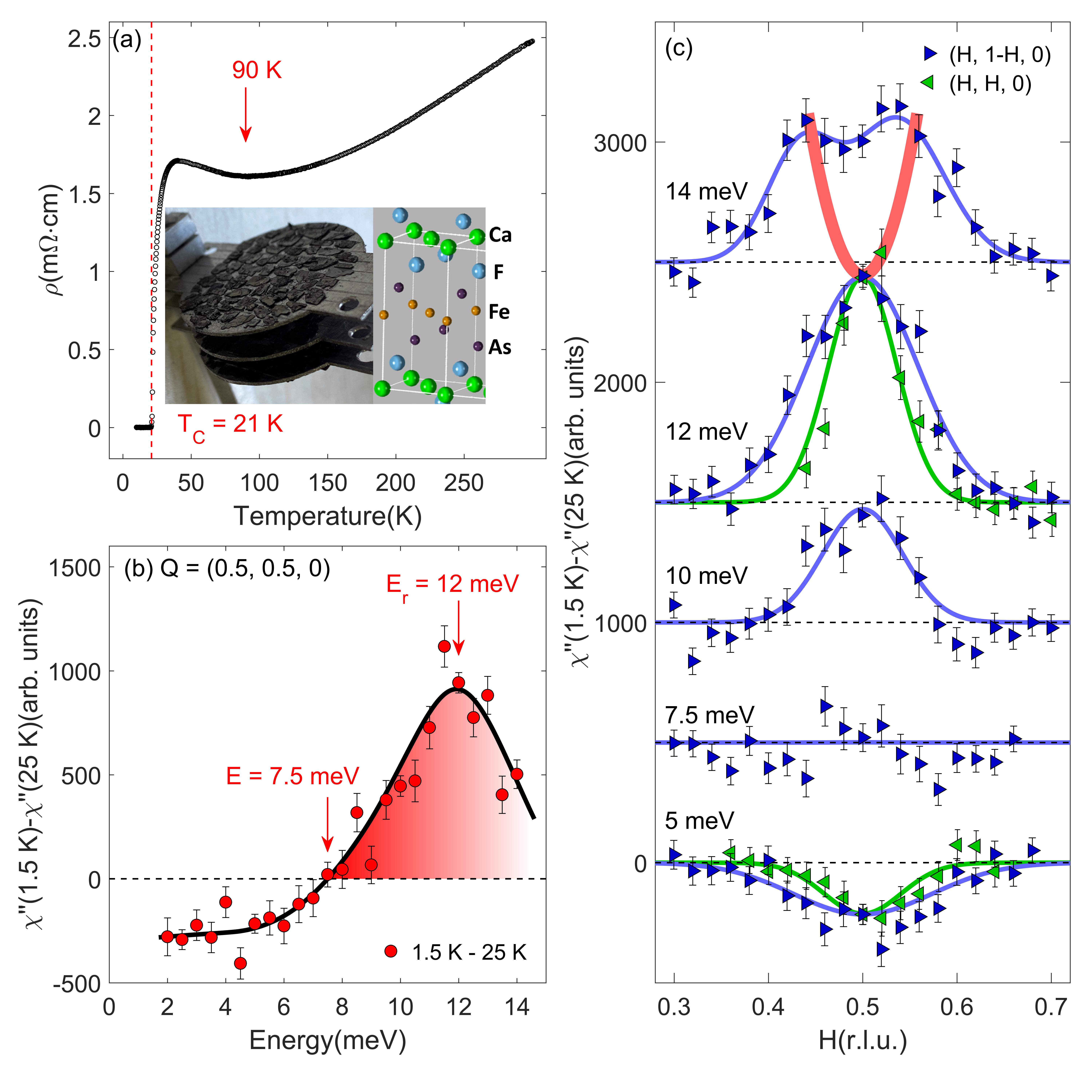}
\caption{\label{Fig1}
(a) Temperature dependence of resistivity, crystal structure and single crystals picture. (b) Temperature difference of imaginary part of the magnetic susceptibility  $\chi^{\prime\prime}$($\mathbf{Q}$,$\omega$) by energy scan between the normal state at $T$ = 25 K and the superconducting state at $T$ = 1.5 K for $\mathbf{Q}_\mathrm{AF}$ = (0.5, 0.5, 0) and $E$ ranging from 2 meV to 14 meV. The black line is guide to the eyes. (c) Temperature difference of imaginary part of the magnetic susceptibility by transverse and longitudinal $\mathbf{Q}$ scan between the normal state at $T$ = 25 K and the superconducting state at $T$ = 1.5 K for $E$ = 5 meV, 7.5 meV, 10 meV, 12 meV and 14 meV respectively. The blue and green lines are fitted by a Gaussian function. The red line is guide to the eyes.
}
\end{figure}

As a prototypal FeAs-1111 system, the parent compound CaFeAsF has a ZrCuSiAs-type tetragonal structure, as shown in the right inset of Fig.~\ref{Fig1}(a). The system undergoes a tetragonal to orthorhombic phase transition at 134 K, whereas antiferromagnetic order further sets in at 114 K. The antiferromagnetic order is completely suppressed in optimally doped CaFe$_{0.88}$Co$_{0.12}$AsF \cite{PhysRevB.79.060504}. Temperature dependence of resistivity for the single crystal CaFe$_{0.88}$Co$_{0.12}$AsF is shown in Fig.~\ref{Fig1}(a). The resistivity as a function of temperature drops down (or falls) gradually with the decrease of temperature in the range above 90 K. \color{black}Below 90 K, the resistivity exhibits a clear upward semiconducting behavior and a sudden drop at $T_\mathrm{c}$ = 21 K. Here we report an INS study on the low-energy spin excitations in the optimally electron-doped CaFe$_{0.88}$Co$_{0.12}$AsF single crystals. We find that the resonance energy $E_\mathrm{r}$ = 12 meV amounts to 6.6 $k_\mathrm{B}$$T_\mathrm{c}$ contrary to 3.7 $k_\mathrm{B}$$T_\mathrm{c}$ ($E_\mathrm{r}$ = 7 meV) in previous INS report on powder sample \cite{Price2013}. Also, the resonance possesses a magnonlike upward dispersion along transverse direction due to the anisotropy of spin-spin correlation length within $ab$ plane in the normal-state.  

Our neutron scattering experiment was carried out on IN8-thermal neutron three-axis spectrometer at Institut Laue-Langevin, Grenoble, France. The study was enabled by a recent breakthrough in the crystal-growth technique for the CaFeAsF compound family, where we utilized temperature oscillations during the growth \cite{MA2022126562}. We used horizontally and vertically focused pyrolytic graphite [PG (002)] monochromator and analyzer with fixed scattered (final) energy  $E_\mathrm{f}$ = 14.7 meV. The high order harmonics from the PG (002) monochromator are suppressed by an oriented PG-filter in the scattered beam. The sample consisted of over 300 pieces of CaFe$_{0.88}$Co$_{0.12}$AsF single crystals with a total mass of about 1.0 g which were grown by the similar temperature oscillating technique [See Supplementary Materials]. As shown in the left inset of  Fig.~\ref{Fig1}(a), these crystals were coaligned within 6$^\circ$ mosaicity in the ($H$, $K$, 0) scattering plane on aluminum plates using a hydrogen-free adhesive. Here and throughout this letter, the wave vector  $\mathbf{Q}$ is expressed in reciprocal lattice units (r.l.u.) as ($H$, $K$, $L$). Using the tetragonal crystal structure, the wave vector, expressed in inverse Angstrom, is $\mathbf{Q}$ = ($(2\pi/a)H$, $(2\pi/a)K$, $(2\pi/c)L$) with lattice parameters $a$ = 3.87 $\mathrm{\AA}$  and $c$ = 8.58 $\mathrm{\AA}$.   \color{black}  Raw data were converted into the imaginary part of the magnetic susceptibility $\chi^{\prime\prime}$($\mathbf{Q}$,$\omega$) after subtracting estimates of the nonmagnetic background and correcting for the Bose population factor [See Supplementary Materials].

Figure~\ref{Fig1}(b) shows the temperature difference of imaginary part of the magnetic susceptibility  $\chi^{\prime\prime}$($\mathbf{Q}$,$\omega$) for CaFe$_{0.88}$Co$_{0.12}$AsF between the normal state at $T$ = 25 K and the SC state at $T$ = 1.5 K with $\mathbf{Q}_\mathrm{AF}$ = (0.5, 0.5, 0) in tetragonal coordinates and energy transfer $E = 2-14$ meV. Below $T_\mathrm{c}$, we observe a suppression of the low energy magnetic response as compared to the normal state below $E$ = 7.5 meV, followed by a net enhancement of the spin supsceptibility at  $E_\mathrm{r}$ = 12 meV. The enhancement of the spin susceptibility in the SC state is a hallmark of the spin resonance excitations. That energy is noticeably larger than the resonance peak which has been previously attributed at 7.5 meV in a powder sample of  CaFe$_{0.88}$Co$_{0.12}$AsF \cite{Price2013} where all crystal orientations are necessarilly averaged out. Clearly, no resonance peak is observed at that energy in our data. Our analysis of the difference of the energy scans at 1.5 K and 25 K at $\mathbf{Q}_\mathrm{AF}$ is further confirmed by a similar measurement at another antiferromagnetic point with larger momentum $\mathbf{Q}_\mathrm{AF'}$ = (1.5, 0.5, 0) [Fig. S1(d)]. The peak is weaker due to the magnetic form factor but it is peaked at the same energy ($E_\mathrm{r}$ = 12 meV). That difference in temperature, which is the usual way to determine the resonance energy \cite{SidisY,InosovNaturePhys2010}, was not performed in the previous study \cite{Price2013}. Instead, the resonance energy  was determined using an extrapolation of the phononic background. As shown by our energy scans at different sample orientations [Fig. S1(a)], the background is clearly not monotonic as assumed in the powder sample study \cite{Price2013}. The measurements on powder sample show two maxima in the raw data at $E$ = 7.5 meV and $E$ = 11 meV, likely due to additional phonon peaks in the neutron spectra. We believe the attribution of a resonance peak energy at 7.5 meV is therefore incorrect and incompatible with our detailed and thorough analysis done on single crystals. \color{black}As we know, the resonant peak energy $E_\mathrm{r}$ in other iron-based superconductors corresponds to 4.9 $k_\mathrm{B}$$T_\mathrm{c}$ in average, which is only slightly different from $E_\mathrm{r}$ $\approx$ 5.3 $k_\mathrm{B}$$T_\mathrm{c}$ for the cuprates and suggests a universal correlation between the superconductivity and the spin excitations in these unconventional superconductors \cite{PhysRevLett.120.137001,SidisY}. Notably, the resonance energy $E_\mathrm{r}$ here amounts to 6.6 $k_\mathrm{B}$$T_\mathrm{c}$ which means that the ratio of $E_\mathrm{r}$/$k_\mathrm{B}$$T_\mathrm{c}$ is the highest value among the previously reported values of $E_\mathrm{r}$/$k_\mathrm{B}$$T_\mathrm{c}$ in iron-based superconductors and might imply a stronger coupling between free electrons and magnetic excitations in this material compared to the other pnictides. 

\begin{figure}
\includegraphics[width=6in]{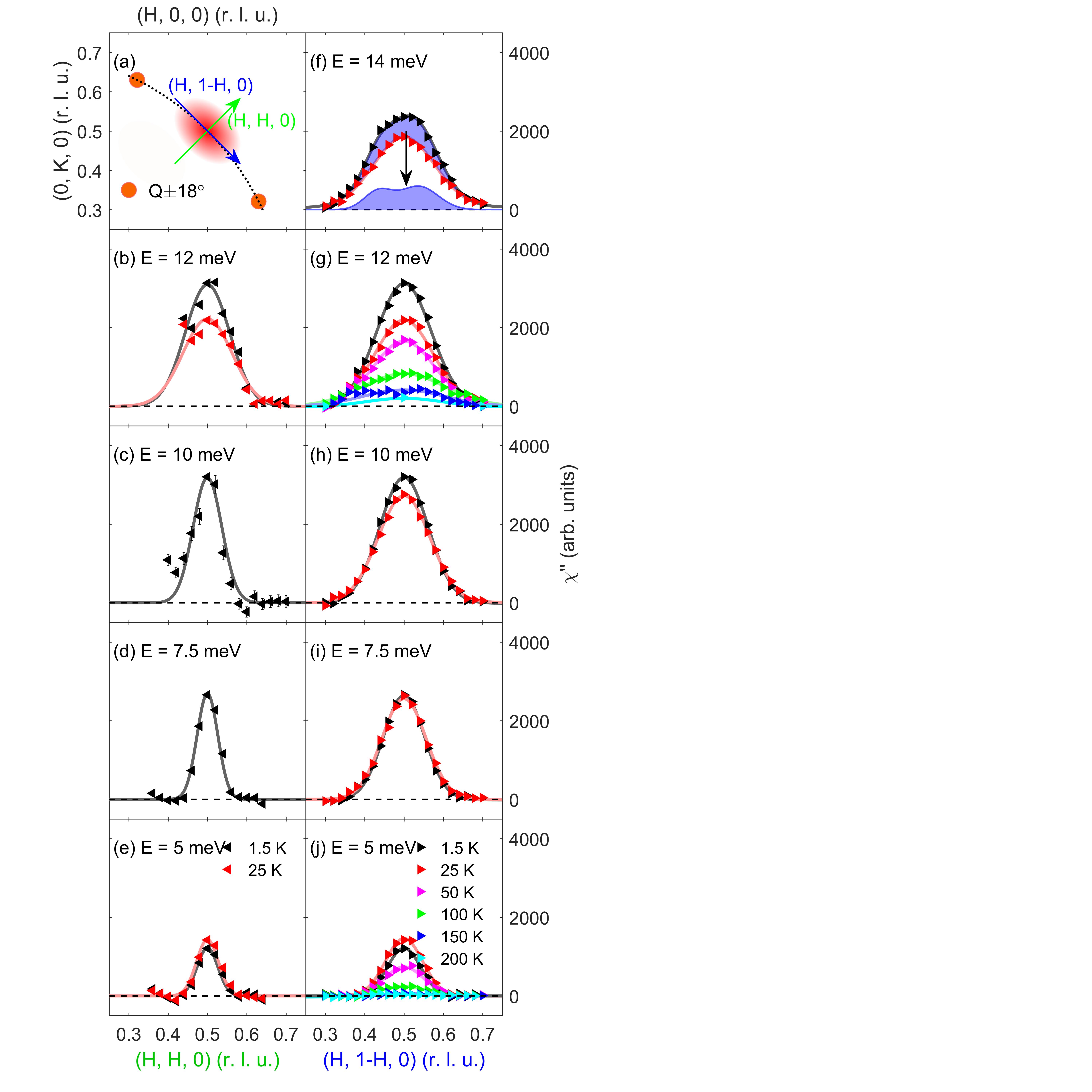}
\caption{\label{Fig2}
(a) Schematic diagram of longitudinal and transverse scan. $\mathbf{Q}$$\pm18$$^\circ$ marked in the Brillouin zone are measured by rotating the sample with $\pm18$$^\circ$ and used to subtract the non-magnetic background. [See Fig. S1 in Supplementary Materials]. \color{black}(b-e) Longitudinal $\mathbf{Q}$ scans of the imaginary part of the magnetic susceptibility  $\chi^{\prime\prime}$($\mathbf{Q}$,$\omega$) at various energy transfers in the normal state at $T$ = 25 K (red triangle) and the superconducting state at $T$ = 1.5 K (black triangle). (f-j) Transverse $\mathbf{Q}$ scans of the imaginary part of the magnetic susceptibility  $\chi^{\prime\prime}$($\mathbf{Q}$,$\omega$) at various energy transfers and temperatures. The solid lines are fitted by a Gaussian function except for the $\mathbf{Q}$ scan at $T$ = 1.5 K and $E$ = 14 meV (f) which is fitted by summing a Gaussian function ($\mathbf{Q}$ scan at  $T$ = 25 K and $E$ = 14 meV) and two symmetric Gaussian peaks. The double peak structure of the $\mathbf{Q}$ scan at $T$ = 1.5 K and $E$ = 14 meV (f) can be clearly shown in the blue shaded area with a flat background. \color{black} The background was estimated by the $\mathbf{Q}$ scans [See Fig. S2-S3 in Supplementary Materials]. 
}
\end{figure}

\color{black}As shown in Fig.~\ref{Fig1}(c), a magnonlike dispersion of spin resonance is confirmed by the temperature difference of constant-energy $\mathbf{Q}$ scans between the susceptibility in the normal and SC states. The resonance disperses upwards away from $\mathbf{Q}_\mathrm{AF}$ = (0.5, 0.5, 0) along the transverse ($H$, 1-$H$, 0) direction. As the energy is increased, the resonance peak broadens and finally splits into two symmetric peaks around $\mathbf{Q}_\mathrm{AF}$ at $E$ = 14 meV indicating the spectral weight progressively moves away from $\mathbf{Q}_\mathrm{AF}$. In contrast, along the longitudinal ($H$, $H$, 0) direction in our scattering geometry, we find that there is no indication of variation of the feature's momentum position versus energy. This observation is consistent with previous work on electron-doped Ba(Fe$_{0.963}$Ni$_{0.037}$)$_{2}$As$_{2}$ where the resonance is found to peak sharply at $\mathbf{Q}_\mathrm{AF}$ along the longitudinal direction, but disperses upwards away from $\mathbf{Q}_\mathrm{AF}$ along the transverse direction \cite{PhysRevLett.110.177002}. Also, there exists a ringlike upward dispersion of spin resonance away from $\mathbf{Q}_\mathrm{AF}$ along both the longitudinal and transverse direction in Ba$_{0.67}$K$_{0.33}$(Fe$_{1-x}$Co$_{x}$)$_{2}$As$_{2}$ \cite{PhysRevB.98.060502}. In contrast, a downward dispersions of spin resonance was observed in KCa$_{2}$Fe$_{4}$As$_{4}$F$_{2}$ superconductor \cite{PhysRevLett.125.117002}, a behavior closely resembling the low branch of the hourglass-type spin resonance in cuprates \cite{Eschrig2006}. Theoretical calculations within the spin-exciton model for iron-based superconductors predict both an upward and a downward dispersion depending on the details of the bands and the symmetry of the superconducting order parameter \cite{PhysRevB.82.134527,PhysRevB.86.094514,PhysRevB.78.140509,PhysRevLett.106.157004}. In a ${s}\mathbf\pm$ superconducting pairing state driven by commensurate short-range spin fluctuations, the resonance mode disperses with increasing energy in an anisotropic pattern broadening most rapidly along the transverse rather than the longitudinal direction and displays an elliptical shape of the spin-resonance mode in $\mathbf{Q}$-space \cite{PhysRevB.82.134527} as shown in the schematic diagram Fig.~\ref{Fig2}(a).\color{black}

\begin{figure}
\includegraphics[width=3.7in]{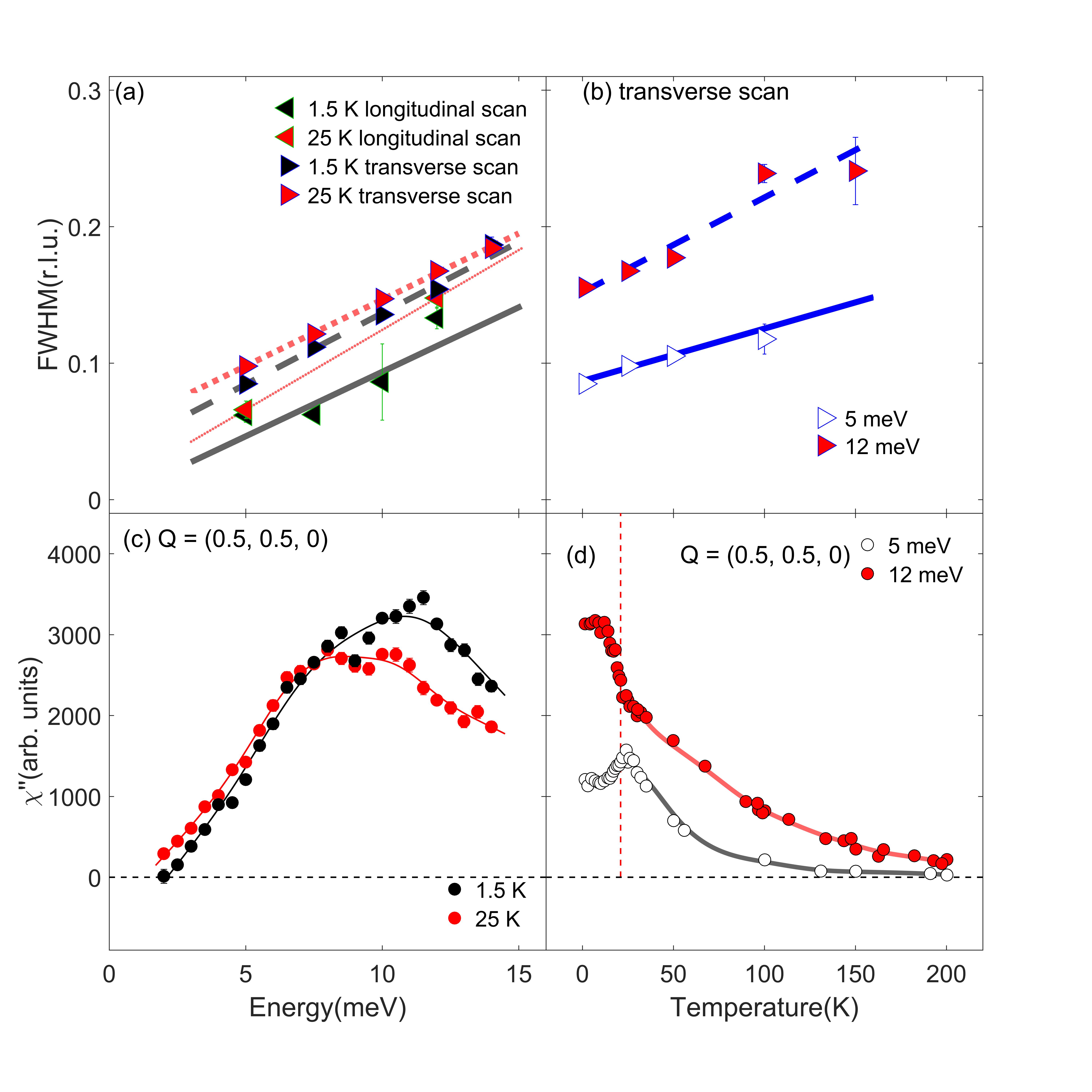}
\caption{\label{Fig3}
(a) Energy dependence of the FWHM of spin excitations at $T$ = 1.5 K and 25 K along longitudinal and transverse directions. (b) Temperature dependence of FWHM of spin excitations at $E$ = 5 meV and 12 meV along transverse direction. (c) Energy dependence of the imaginary part of the magnetic susceptibility  $\chi^{\prime\prime}$($\mathbf{Q}$,$\omega$) at $T$ = 1.5 K and 25 K and $\mathbf{Q}_\mathrm{AF}$ = (0.5, 0.5, 0). The background was estimated by the measurement with a 18$^\circ$ deviation of A3 angle. (See Fig. S1 in Supplementary Materials). (d) Temperature dependence of the imaginary part of the magnetic susceptibility  $\chi^{\prime\prime}$($\mathbf{Q}$,$\omega$) at E = 5 meV and 12 meV and $\mathbf{Q}_\mathrm{AF}$ = (0.5, 0.5, 0). The background was estimated by the $\mathbf{Q}$-scans at various temperatures [See Fig. S3 in Supplementary Materials]. All the lines are guide to the eyes.
}
\end{figure}
 \color{black}
The $\mathbf{Q}$-space anisotropy of the spin resonance in the SC state may be associated with the normal state anisotropic scattering pattern found locally around $\mathbf{Q}_\mathrm{AF}$, which is a consequence of the fermiology, the shape of the electron and hole like pockets involved in the particle-hole excitation process. The close relationship between the upward spin resonance dispersion and normal state spin fluctuations anisotropy can be illustrated from constant-energy $\mathbf{Q}$ scans in the normal and superconducting states in the vicinity of $\mathbf{Q}_\mathrm{AF}$ = (0.5, 0.5, 0), focusing on the longitudinal and transverse direction as displayed in the schematic diagram Fig.~\ref{Fig2}(a). Background scattering measured by rotating the sample by $\pm18$$^\circ$ has been removed in order to display the genuine magnetic signals. \color{black} Figure~\ref{Fig2}(b)-(j) shows the imaginary part of the magnetic susceptibility  $\chi^{\prime\prime}$($\mathbf{Q}$,$\omega$) for CaFe$_{0.88}$Co$_{0.12}$AsF at various temperature and energy transfer $E$ along longitudinal and transverse $\mathbf{Q}$-scans respectively. With increasing energy the widths of the respective peaks at the corresponding temperatures and scan directions become broader. Notably, the longitudinal scans show much steeper lineshape than transverse scans. The signal's transverse momentum profile is further found to become broader with increasing temperature [Fig.~\ref{Fig2}(g, j)] along with a decrease in the peak amplitude.

To clearly illustrate the $\mathbf{Q}$-space anisotropy, we compare the full-width-at-half-maximum (FWHM) between transverse and longitudinal scans in Fig.~\ref{Fig3}(a), in both the SC and the normal states. The FWHM of peak obtained by Gaussian fitting is associated with the spin-spin correlation length. The FWHM of transverse scan is apparently larger than that of longitudinal scan. The difference in linewidth of the longitudinal (narrower) and transverse (broader) scans is not due to the instrument resolution but from the intrinsic anisotropy in momentum $\mathbf{Q}$-space of the spin fluctuations. Actually, the instrumental resolution is roughly isotropic in the experiment. Therefore, the double peak structure does not show up in longitudinal scans as the peak is sharper along that direction compared to the transverse scans. \color{black} Within the probed energy range ($E = 5-14$ meV), the widths of spin excitations increase monotonically with increasing energy. Figure~\ref{Fig3}(b) shows the temperature dependence of the spin excitation widths along transverse direction at $E$ = 5 and 12 meV. The temperature dependent broadening indicates a gradual reduction in the spin-correlation length with temperature which behavior has also been observed in LaFeAsO, Co/Ni/K doped BaFe$_{2}$As$_{2}$ \cite{PhysRevLett.114.057001,PhysRevLett.117.227003,WangMNC} and references therein.

\color{black}
Figure~\ref{Fig3}(c) shows the energy dependence of the imaginary part of the magnetic susceptibility  $\chi^{\prime\prime}$($\mathbf{Q}$,$\omega$) at $T$ = 1.5 K and 25 K and $\mathbf{Q}_\mathrm{AF}$ = (0.5, 0.5, 0). The spin susceptibility develops with increasing energy, shows a hump around 10 meV and then decreases gradually as the FWHM broadening with energy [Fig.~\ref{Fig3}(a)]. Figure~\ref{Fig3}(d) displays the detailed temperature dependence of the magnetic susceptibility $\chi^{\prime\prime}$($\mathbf{Q}$,$\omega$) for $E$ = 5 and 12 meV and $\mathbf{Q}_\mathrm{AF}$ = (0.5, 0.5, 0), which shows pronounced anomalies around $T_\mathrm{c}$. The observation is consistent with a redistribution of the scattering intensity, which may also involve momenta slightly away from $\mathbf{Q}_\mathrm{AF}$ as shown by the gradual increase of the FWHM in Fig.~\ref{Fig3}(b). 

Given the intrinsic nature of the anisotropy of spin fluctuations in  CaFe$_{0.88}$Co$_{0.12}$AsF, one arrives to the conclusion of a close relationship between the upward dispersion of the spin resonance and the anisotropic spin fluctuations in the normal state. The fully gapped ${s}\mathbf\pm$ wave superconductivity can be driven by the short-range spin fluctuations which generally exhibit an anisotropy in momentum space with $\mathbf{Q}$ width larger along the direction transverse to $\mathbf{Q}_\mathrm{AF}$ than along the longitudinal direction and can be understood by examining the intra-orbital scattering processes in systems away from perfect nesting \cite{PhysRevB.82.134527}. The same type of anisotropy in the SC state gives rise to an elliptical shape of the spin resonant mode which disperses with increasing energy in the pattern broadening more rapidly along the transverse than along the longitudinal direction \cite{PhysRevB.82.134527}.  A more elliptically shaped $\mathbf{Q}$ image of the spin resonant mode can be produced by the intrinsic anisotropy in momentum space associated with anisotropic dispersion of the resonant mode \cite{PhysRevB.82.134527}.\color{black}

In conclusion, we have observed a sharp superconducting spin resonance at resonance energy $E_\mathrm{r}$ = 12 meV. The magnonlike dispersion of spin resonance can be attributed to the $\mathbf{Q}$-space anisotropy of spin excitations in the normal state of optimally doped  CaFe$_{0.88}$Co$_{0.12}$AsF, and it indicates sign-reversed  ${s}\mathbf\pm$ wave pairing possibly mediated by magnetic fluctuations.

\begin{acknowledgments}
The work was supported by the National Natural Science Foundation of China (Grant No. 12004418), the National Key Research and Development of China (Grant No. 2018YFA0704200, 2022YFA1602803) and the Strategic Priority Research Program of Chinese Academy of Sciences (Grant No. XDB25000000).
\end{acknowledgments}

\preprint{Preprint}

%


\begin{thebibliography}{34}%
\makeatletter
\providecommand \@ifxundefined [1]{%
 \@ifx{#1\undefined}
}%
\providecommand \@ifnum [1]{%
 \ifnum #1\expandafter \@firstoftwo
 \else \expandafter \@secondoftwo
 \fi
}%
\providecommand \@ifx [1]{%
 \ifx #1\expandafter \@firstoftwo
 \else \expandafter \@secondoftwo
 \fi
}%
\providecommand \natexlab [1]{#1}%
\providecommand \enquote  [1]{``#1''}%
\providecommand \bibnamefont  [1]{#1}%
\providecommand \bibfnamefont [1]{#1}%
\providecommand \citenamefont [1]{#1}%
\providecommand \href@noop [0]{\@secondoftwo}%
\providecommand \href [0]{\begingroup \@sanitize@url \@href}%
\providecommand \@href[1]{\@@startlink{#1}\@@href}%
\providecommand \@@href[1]{\endgroup#1\@@endlink}%
\providecommand \@sanitize@url [0]{\catcode `\\12\catcode `\$12\catcode
  `\&12\catcode `\#12\catcode `\^12\catcode `\_12\catcode `\%12\relax}%
\providecommand \@@startlink[1]{}%
\providecommand \@@endlink[0]{}%
\providecommand \url  [0]{\begingroup\@sanitize@url \@url }%
\providecommand \@url [1]{\endgroup\@href {#1}{\urlprefix }}%
\providecommand \urlprefix  [0]{URL }%
\providecommand \Eprint [0]{\href }%
\providecommand \doibase [0]{http://dx.doi.org/}%
\providecommand \selectlanguage [0]{\@gobble}%
\providecommand \bibinfo  [0]{\@secondoftwo}%
\providecommand \bibfield  [0]{\@secondoftwo}%
\providecommand \translation [1]{[#1]}%
\providecommand \BibitemOpen [0]{}%
\providecommand \bibitemStop [0]{}%
\providecommand \bibitemNoStop [0]{.\EOS\space}%
\providecommand \EOS [0]{\spacefactor3000\relax}%
\providecommand \BibitemShut  [1]{\csname bibitem#1\endcsname}%
\let\auto@bib@innerbib\@empty
\bibitem [{\citenamefont {Rossat-Mignod}\ \emph {et~al.}(1991)\citenamefont
  {Rossat-Mignod}, \citenamefont {Regnault}, \citenamefont {Vettier},
  \citenamefont {Bourges}, \citenamefont {Burlet}, \citenamefont {Bossy},
  \citenamefont {Henry},\ and\ \citenamefont {Lapertot}}]{RossatMignod}%
  \BibitemOpen
  \bibfield  {author} {\bibinfo {author} {\bibfnamefont {J.}~\bibnamefont
  {Rossat-Mignod}}, \bibinfo {author} {\bibfnamefont {L.~P.}\ \bibnamefont
  {Regnault}}, \bibinfo {author} {\bibfnamefont {C.}~\bibnamefont {Vettier}},
  \bibinfo {author} {\bibfnamefont {P.}~\bibnamefont {Bourges}}, \bibinfo
  {author} {\bibfnamefont {P.}~\bibnamefont {Burlet}}, \bibinfo {author}
  {\bibfnamefont {J.}~\bibnamefont {Bossy}}, \bibinfo {author} {\bibfnamefont
  {J.~Y.}\ \bibnamefont {Henry}}, \ and\ \bibinfo {author} {\bibfnamefont
  {G.}~\bibnamefont {Lapertot}},\ }\href@noop {} {\bibfield  {journal}
  {\bibinfo  {journal} {Physica C-superconductivity and Its Applications}\ ,\
  \bibinfo {pages} {86}} (\bibinfo {year} {1991})}\BibitemShut {NoStop}%
\bibitem [{\citenamefont {Mook}\ \emph {et~al.}(1993)\citenamefont {Mook},
  \citenamefont {Yethiraj}, \citenamefont {Aeppli}, \citenamefont {Mason},\
  and\ \citenamefont {Armstrong}}]{PhysRevLett.70.3490}%
  \BibitemOpen
  \bibfield  {author} {\bibinfo {author} {\bibfnamefont {H.~A.}\ \bibnamefont
  {Mook}}, \bibinfo {author} {\bibfnamefont {M.}~\bibnamefont {Yethiraj}},
  \bibinfo {author} {\bibfnamefont {G.}~\bibnamefont {Aeppli}}, \bibinfo
  {author} {\bibfnamefont {T.~E.}\ \bibnamefont {Mason}}, \ and\ \bibinfo
  {author} {\bibfnamefont {T.}~\bibnamefont {Armstrong}},\ }\href {\doibase
  10.1103/PhysRevLett.70.3490} {\bibfield  {journal} {\bibinfo  {journal}
  {Phys. Rev. Lett.}\ }\textbf {\bibinfo {volume} {70}},\ \bibinfo {pages}
  {3490} (\bibinfo {year} {1993})}\BibitemShut {NoStop}%
\bibitem [{\citenamefont {Sato}\ \emph {et~al.}(2001)\citenamefont {Sato},
  \citenamefont {N.~Aso}, \citenamefont {R.~Shiina}, \citenamefont
  {Varelogiannis}, \citenamefont {Geibel}, \citenamefont {Steglich},
  \citenamefont {Fulde},\ and\ \citenamefont {Komatsubara}}]{SatoNature2001}%
  \BibitemOpen
  \bibfield  {author} {\bibinfo {author} {\bibfnamefont {N.~K.}\ \bibnamefont
  {Sato}}, \bibinfo {author} {\bibfnamefont {K.~M.}\ \bibnamefont {N.~Aso}},
  \bibinfo {author} {\bibfnamefont {P.~T.}\ \bibnamefont {R.~Shiina}}, \bibinfo
  {author} {\bibfnamefont {G.}~\bibnamefont {Varelogiannis}}, \bibinfo {author}
  {\bibfnamefont {C.}~\bibnamefont {Geibel}}, \bibinfo {author} {\bibfnamefont
  {F.}~\bibnamefont {Steglich}}, \bibinfo {author} {\bibfnamefont
  {P.}~\bibnamefont {Fulde}}, \ and\ \bibinfo {author} {\bibfnamefont
  {T.}~\bibnamefont {Komatsubara}},\ }\href {\doibase
  https://doi.org/10.1038/35066519} {\bibfield  {journal} {\bibinfo  {journal}
  {Nature}\ }\textbf {\bibinfo {volume} {410}},\ \bibinfo {pages} {340}
  (\bibinfo {year} {2001})}\BibitemShut {NoStop}%
\bibitem [{\citenamefont {Stock}\ \emph {et~al.}(2008)\citenamefont {Stock},
  \citenamefont {Broholm}, \citenamefont {Hudis}, \citenamefont {Kang},\ and\
  \citenamefont {Petrovic}}]{PhysRevLett.100.087001}%
  \BibitemOpen
  \bibfield  {author} {\bibinfo {author} {\bibfnamefont {C.}~\bibnamefont
  {Stock}}, \bibinfo {author} {\bibfnamefont {C.}~\bibnamefont {Broholm}},
  \bibinfo {author} {\bibfnamefont {J.}~\bibnamefont {Hudis}}, \bibinfo
  {author} {\bibfnamefont {H.~J.}\ \bibnamefont {Kang}}, \ and\ \bibinfo
  {author} {\bibfnamefont {C.}~\bibnamefont {Petrovic}},\ }\href {\doibase
  10.1103/PhysRevLett.100.087001} {\bibfield  {journal} {\bibinfo  {journal}
  {Phys. Rev. Lett.}\ }\textbf {\bibinfo {volume} {100}},\ \bibinfo {pages}
  {087001} (\bibinfo {year} {2008})}\BibitemShut {NoStop}%
\bibitem [{\citenamefont {Christianson}\ \emph {et~al.}(2008)\citenamefont
  {Christianson}, \citenamefont {Goremychkin}, \citenamefont {Osborn},
  \citenamefont {Rosenkranz}, \citenamefont {Lumsden}, \citenamefont
  {Malliakas}, \citenamefont {Todorov}, \citenamefont {Claus}, \citenamefont
  {Chung}, \citenamefont {Kanatzidis}, \citenamefont {Bewley},\ and\
  \citenamefont {Guidi}}]{ChristiansonNature2008}%
  \BibitemOpen
  \bibfield  {author} {\bibinfo {author} {\bibfnamefont {A.~D.}\ \bibnamefont
  {Christianson}}, \bibinfo {author} {\bibfnamefont {E.~A.}\ \bibnamefont
  {Goremychkin}}, \bibinfo {author} {\bibfnamefont {R.}~\bibnamefont {Osborn}},
  \bibinfo {author} {\bibfnamefont {S.}~\bibnamefont {Rosenkranz}}, \bibinfo
  {author} {\bibfnamefont {M.~D.}\ \bibnamefont {Lumsden}}, \bibinfo {author}
  {\bibfnamefont {C.~D.}\ \bibnamefont {Malliakas}}, \bibinfo {author}
  {\bibfnamefont {I.~S.}\ \bibnamefont {Todorov}}, \bibinfo {author}
  {\bibfnamefont {H.}~\bibnamefont {Claus}}, \bibinfo {author} {\bibfnamefont
  {D.~Y.}\ \bibnamefont {Chung}}, \bibinfo {author} {\bibfnamefont {M.~G.}\
  \bibnamefont {Kanatzidis}}, \bibinfo {author} {\bibfnamefont {R.~I.}\
  \bibnamefont {Bewley}}, \ and\ \bibinfo {author} {\bibfnamefont
  {T.}~\bibnamefont {Guidi}},\ }\href {\doibase
  https://doi.org/10.1038/nature07625} {\bibfield  {journal} {\bibinfo
  {journal} {Nature}\ }\textbf {\bibinfo {volume} {456}},\ \bibinfo {pages}
  {930} (\bibinfo {year} {2008})}\BibitemShut {NoStop}%
\bibitem [{\citenamefont {Liu}\ \emph {et~al.}(2010)\citenamefont {Liu},
  \citenamefont {Hu}, \citenamefont {Qian}, \citenamefont {Fobes},
  \citenamefont {Mao}, \citenamefont {Bao}, \citenamefont {Reehuis},
  \citenamefont {Kimber}, \citenamefont {Prokes}, \citenamefont {Matas},
  \citenamefont {Argyriou}, \citenamefont {Hiess}, \citenamefont {Rotaru},
  \citenamefont {Pham}, \citenamefont {Spinu}, \citenamefont {Qiu},
  \citenamefont {Thampy}, \citenamefont {Savici}, \citenamefont {Rodriguez},\
  and\ \citenamefont {Broholm}}]{LiuNatureMater2010}%
  \BibitemOpen
  \bibfield  {author} {\bibinfo {author} {\bibfnamefont {T.~J.}\ \bibnamefont
  {Liu}}, \bibinfo {author} {\bibfnamefont {J.}~\bibnamefont {Hu}}, \bibinfo
  {author} {\bibfnamefont {B.}~\bibnamefont {Qian}}, \bibinfo {author}
  {\bibfnamefont {D.}~\bibnamefont {Fobes}}, \bibinfo {author} {\bibfnamefont
  {Z.~Q.}\ \bibnamefont {Mao}}, \bibinfo {author} {\bibfnamefont
  {W.}~\bibnamefont {Bao}}, \bibinfo {author} {\bibfnamefont {M.}~\bibnamefont
  {Reehuis}}, \bibinfo {author} {\bibfnamefont {S.~A.~J.}\ \bibnamefont
  {Kimber}}, \bibinfo {author} {\bibfnamefont {K.}~\bibnamefont {Prokes}},
  \bibinfo {author} {\bibfnamefont {S.}~\bibnamefont {Matas}}, \bibinfo
  {author} {\bibfnamefont {D.~N.}\ \bibnamefont {Argyriou}}, \bibinfo {author}
  {\bibfnamefont {A.}~\bibnamefont {Hiess}}, \bibinfo {author} {\bibfnamefont
  {A.}~\bibnamefont {Rotaru}}, \bibinfo {author} {\bibfnamefont
  {H.}~\bibnamefont {Pham}}, \bibinfo {author} {\bibfnamefont {L.}~\bibnamefont
  {Spinu}}, \bibinfo {author} {\bibfnamefont {Y.}~\bibnamefont {Qiu}}, \bibinfo
  {author} {\bibfnamefont {V.}~\bibnamefont {Thampy}}, \bibinfo {author}
  {\bibfnamefont {A.~T.}\ \bibnamefont {Savici}}, \bibinfo {author}
  {\bibfnamefont {J.~A.}\ \bibnamefont {Rodriguez}}, \ and\ \bibinfo {author}
  {\bibfnamefont {C.}~\bibnamefont {Broholm}},\ }\href {\doibase
  https://doi.org/10.1038/nmat2800} {\bibfield  {journal} {\bibinfo  {journal}
  {Nat. Mater.}\ }\textbf {\bibinfo {volume} {9}},\ \bibinfo {pages} {718}
  (\bibinfo {year} {2010})}\BibitemShut {NoStop}%
\bibitem [{\citenamefont {Eschrig}(2006)}]{Eschrig2006}%
  \BibitemOpen
  \bibfield  {author} {\bibinfo {author} {\bibfnamefont {M.}~\bibnamefont
  {Eschrig}},\ }\href {\doibase https://doi.org/10.1080/00018730600645636}
  {\bibfield  {journal} {\bibinfo  {journal} {Adv. Phys.}\ }\textbf {\bibinfo
  {volume} {55}},\ \bibinfo {pages} {47} (\bibinfo {year} {2006})}\BibitemShut
  {NoStop}%
\bibitem [{\citenamefont {Yu}\ \emph {et~al.}(2009)\citenamefont {Yu},
  \citenamefont {Li}, \citenamefont {Motoyama},\ and\ \citenamefont
  {Greven}}]{YuNaturePhys2009}%
  \BibitemOpen
  \bibfield  {author} {\bibinfo {author} {\bibfnamefont {G.}~\bibnamefont
  {Yu}}, \bibinfo {author} {\bibfnamefont {Y.}~\bibnamefont {Li}}, \bibinfo
  {author} {\bibfnamefont {E.~M.}\ \bibnamefont {Motoyama}}, \ and\ \bibinfo
  {author} {\bibfnamefont {M.~A.}\ \bibnamefont {Greven}},\ }\href {\doibase
  https://doi.org/10.1038/nphys1426} {\bibfield  {journal} {\bibinfo  {journal}
  {Nat.Phys.}\ }\textbf {\bibinfo {volume} {5}},\ \bibinfo {pages} {873}
  (\bibinfo {year} {2009})}\BibitemShut {NoStop}%
\bibitem [{\citenamefont {Inosov}\ \emph {et~al.}(2011)\citenamefont {Inosov},
  \citenamefont {Park}, \citenamefont {Charnukha}, \citenamefont {Li},
  \citenamefont {Boris}, \citenamefont {Keimer},\ and\ \citenamefont
  {Hinkov}}]{PhysRevB.83.214520}%
  \BibitemOpen
  \bibfield  {author} {\bibinfo {author} {\bibfnamefont {D.~S.}\ \bibnamefont
  {Inosov}}, \bibinfo {author} {\bibfnamefont {J.~T.}\ \bibnamefont {Park}},
  \bibinfo {author} {\bibfnamefont {A.}~\bibnamefont {Charnukha}}, \bibinfo
  {author} {\bibfnamefont {Y.}~\bibnamefont {Li}}, \bibinfo {author}
  {\bibfnamefont {A.~V.}\ \bibnamefont {Boris}}, \bibinfo {author}
  {\bibfnamefont {B.}~\bibnamefont {Keimer}}, \ and\ \bibinfo {author}
  {\bibfnamefont {V.}~\bibnamefont {Hinkov}},\ }\href {\doibase
  10.1103/PhysRevB.83.214520} {\bibfield  {journal} {\bibinfo  {journal} {Phys.
  Rev. B}\ }\textbf {\bibinfo {volume} {83}},\ \bibinfo {pages} {214520}
  (\bibinfo {year} {2011})}\BibitemShut {NoStop}%
\bibitem [{\citenamefont {Sidis}\ \emph {et~al.}(2007)\citenamefont {Sidis},
  \citenamefont {Pailh\`es}, \citenamefont {Hinkov}, \citenamefont {Fauqu\'e},
  \citenamefont {Ulrich}, \citenamefont {Capogna}, \citenamefont {Ivanov},
  \citenamefont {Regnault}, \citenamefont {Keimer},\ and\ \citenamefont
  {Bourges}}]{SIDIS2007745}%
  \BibitemOpen
  \bibfield  {author} {\bibinfo {author} {\bibfnamefont {Y.}~\bibnamefont
  {Sidis}}, \bibinfo {author} {\bibfnamefont {S.}~\bibnamefont {Pailh\`es}},
  \bibinfo {author} {\bibfnamefont {V.}~\bibnamefont {Hinkov}}, \bibinfo
  {author} {\bibfnamefont {B.}~\bibnamefont {Fauqu\'e}}, \bibinfo {author}
  {\bibfnamefont {C.}~\bibnamefont {Ulrich}}, \bibinfo {author} {\bibfnamefont
  {L.}~\bibnamefont {Capogna}}, \bibinfo {author} {\bibfnamefont
  {A.}~\bibnamefont {Ivanov}}, \bibinfo {author} {\bibfnamefont {L.-P.}\
  \bibnamefont {Regnault}}, \bibinfo {author} {\bibfnamefont {B.}~\bibnamefont
  {Keimer}}, \ and\ \bibinfo {author} {\bibfnamefont {P.}~\bibnamefont
  {Bourges}},\ }\href {\doibase https://doi.org/10.1016/j.crhy.2007.07.003}
  {\bibfield  {journal} {\bibinfo  {journal} {Comptes Rendus Physique}\
  }\textbf {\bibinfo {volume} {8}},\ \bibinfo {pages} {745} (\bibinfo {year}
  {2007})}\BibitemShut {NoStop}%
\bibitem [{\citenamefont {Maier}\ and\ \citenamefont
  {Scalapino}(2008)}]{PhysRevB.78.020514}%
  \BibitemOpen
  \bibfield  {author} {\bibinfo {author} {\bibfnamefont {T.~A.}\ \bibnamefont
  {Maier}}\ and\ \bibinfo {author} {\bibfnamefont {D.~J.}\ \bibnamefont
  {Scalapino}},\ }\href {\doibase 10.1103/PhysRevB.78.020514} {\bibfield
  {journal} {\bibinfo  {journal} {Phys. Rev. B}\ }\textbf {\bibinfo {volume}
  {78}},\ \bibinfo {pages} {020514} (\bibinfo {year} {2008})}\BibitemShut
  {NoStop}%
\bibitem [{\citenamefont {Parish}\ \emph {et~al.}(2008)\citenamefont {Parish},
  \citenamefont {Hu},\ and\ \citenamefont {Bernevig}}]{PhysRevB.78.144514}%
  \BibitemOpen
  \bibfield  {author} {\bibinfo {author} {\bibfnamefont {M.~M.}\ \bibnamefont
  {Parish}}, \bibinfo {author} {\bibfnamefont {J.}~\bibnamefont {Hu}}, \ and\
  \bibinfo {author} {\bibfnamefont {B.~A.}\ \bibnamefont {Bernevig}},\ }\href
  {\doibase 10.1103/PhysRevB.78.144514} {\bibfield  {journal} {\bibinfo
  {journal} {Phys. Rev. B}\ }\textbf {\bibinfo {volume} {78}},\ \bibinfo
  {pages} {144514} (\bibinfo {year} {2008})}\BibitemShut {NoStop}%
\bibitem [{\citenamefont {Korshunov}\ and\ \citenamefont
  {Eremin}(2008)}]{PhysRevB.78.140509}%
  \BibitemOpen
  \bibfield  {author} {\bibinfo {author} {\bibfnamefont {M.~M.}\ \bibnamefont
  {Korshunov}}\ and\ \bibinfo {author} {\bibfnamefont {I.}~\bibnamefont
  {Eremin}},\ }\href {\doibase 10.1103/PhysRevB.78.140509} {\bibfield
  {journal} {\bibinfo  {journal} {Phys. Rev. B}\ }\textbf {\bibinfo {volume}
  {78}},\ \bibinfo {pages} {140509} (\bibinfo {year} {2008})}\BibitemShut
  {NoStop}%
\bibitem [{\citenamefont {Wang}\ and\ \citenamefont {Lee}(2011)}]{Wang2011Sci}%
  \BibitemOpen
  \bibfield  {author} {\bibinfo {author} {\bibfnamefont {F.}~\bibnamefont
  {Wang}}\ and\ \bibinfo {author} {\bibfnamefont {D.-H.}\ \bibnamefont {Lee}},\
  }\href {\doibase 10.1126/science.1200182} {\bibfield  {journal} {\bibinfo
  {journal} {Science}\ }\textbf {\bibinfo {volume} {332}},\ \bibinfo {pages}
  {200} (\bibinfo {year} {2011})}\BibitemShut {NoStop}%
\bibitem [{\citenamefont {Hirschfeld}\ \emph {et~al.}(2011)\citenamefont
  {Hirschfeld}, \citenamefont {Korshunov},\ and\ \citenamefont
  {Mazin}}]{Hirschfeld_2011}%
  \BibitemOpen
  \bibfield  {author} {\bibinfo {author} {\bibfnamefont {P.~J.}\ \bibnamefont
  {Hirschfeld}}, \bibinfo {author} {\bibfnamefont {M.~M.}\ \bibnamefont
  {Korshunov}}, \ and\ \bibinfo {author} {\bibfnamefont {I.~I.}\ \bibnamefont
  {Mazin}},\ }\href {\doibase 10.1088/0034-4885/74/12/124508} {\bibfield
  {journal} {\bibinfo  {journal} {Reports on Progress in Physics}\ }\textbf
  {\bibinfo {volume} {74}},\ \bibinfo {pages} {124508} (\bibinfo {year}
  {2011})}\BibitemShut {NoStop}%
\bibitem [{\citenamefont {Seo}\ \emph {et~al.}(2008)\citenamefont {Seo},
  \citenamefont {Bernevig},\ and\ \citenamefont {Hu}}]{PhysRevLett.101.206404}%
  \BibitemOpen
  \bibfield  {author} {\bibinfo {author} {\bibfnamefont {K.}~\bibnamefont
  {Seo}}, \bibinfo {author} {\bibfnamefont {B.~A.}\ \bibnamefont {Bernevig}}, \
  and\ \bibinfo {author} {\bibfnamefont {J.}~\bibnamefont {Hu}},\ }\href
  {\doibase 10.1103/PhysRevLett.101.206404} {\bibfield  {journal} {\bibinfo
  {journal} {Phys. Rev. Lett.}\ }\textbf {\bibinfo {volume} {101}},\ \bibinfo
  {pages} {206404} (\bibinfo {year} {2008})}\BibitemShut {NoStop}%
\bibitem [{\citenamefont {Mazin}\ and\ \citenamefont
  {Schmalian}(2009)}]{MAZIN2009614}%
  \BibitemOpen
  \bibfield  {author} {\bibinfo {author} {\bibfnamefont {I.}~\bibnamefont
  {Mazin}}\ and\ \bibinfo {author} {\bibfnamefont {J.}~\bibnamefont
  {Schmalian}},\ }\href {\doibase https://doi.org/10.1016/j.physc.2009.03.019}
  {\bibfield  {journal} {\bibinfo  {journal} {Physica C: Superconductivity}\
  }\textbf {\bibinfo {volume} {469}},\ \bibinfo {pages} {614} (\bibinfo {year}
  {2009})}\BibitemShut {NoStop}%
\bibitem [{\citenamefont {Chubukov}\ \emph {et~al.}(2008)\citenamefont
  {Chubukov}, \citenamefont {Efremov},\ and\ \citenamefont
  {Eremin}}]{PhysRevB.78.134512}%
  \BibitemOpen
  \bibfield  {author} {\bibinfo {author} {\bibfnamefont {A.~V.}\ \bibnamefont
  {Chubukov}}, \bibinfo {author} {\bibfnamefont {D.~V.}\ \bibnamefont
  {Efremov}}, \ and\ \bibinfo {author} {\bibfnamefont {I.}~\bibnamefont
  {Eremin}},\ }\href {\doibase 10.1103/PhysRevB.78.134512} {\bibfield
  {journal} {\bibinfo  {journal} {Phys. Rev. B}\ }\textbf {\bibinfo {volume}
  {78}},\ \bibinfo {pages} {134512} (\bibinfo {year} {2008})}\BibitemShut
  {NoStop}%
\bibitem [{\citenamefont {Richard}\ \emph {et~al.}(2011)\citenamefont
  {Richard}, \citenamefont {Sato}, \citenamefont {Nakayama}, \citenamefont
  {Takahashi},\ and\ \citenamefont {Ding}}]{Richard_2011}%
  \BibitemOpen
  \bibfield  {author} {\bibinfo {author} {\bibfnamefont {P.}~\bibnamefont
  {Richard}}, \bibinfo {author} {\bibfnamefont {T.}~\bibnamefont {Sato}},
  \bibinfo {author} {\bibfnamefont {K.}~\bibnamefont {Nakayama}}, \bibinfo
  {author} {\bibfnamefont {T.}~\bibnamefont {Takahashi}}, \ and\ \bibinfo
  {author} {\bibfnamefont {H.}~\bibnamefont {Ding}},\ }\href {\doibase
  10.1088/0034-4885/74/12/124512} {\bibfield  {journal} {\bibinfo  {journal}
  {Reports on Progress in Physics}\ }\textbf {\bibinfo {volume} {74}},\
  \bibinfo {pages} {124512} (\bibinfo {year} {2011})}\BibitemShut {NoStop}%
\bibitem [{\citenamefont {Zhang}\ \emph {et~al.}(2018)\citenamefont {Zhang},
  \citenamefont {Wang}, \citenamefont {Maier}, \citenamefont {Wang},
  \citenamefont {Stone}, \citenamefont {Chi}, \citenamefont {Winn},\ and\
  \citenamefont {Dai}}]{PhysRevB.98.060502}%
  \BibitemOpen
  \bibfield  {author} {\bibinfo {author} {\bibfnamefont {R.}~\bibnamefont
  {Zhang}}, \bibinfo {author} {\bibfnamefont {W.}~\bibnamefont {Wang}},
  \bibinfo {author} {\bibfnamefont {T.~A.}\ \bibnamefont {Maier}}, \bibinfo
  {author} {\bibfnamefont {M.}~\bibnamefont {Wang}}, \bibinfo {author}
  {\bibfnamefont {M.~B.}\ \bibnamefont {Stone}}, \bibinfo {author}
  {\bibfnamefont {S.}~\bibnamefont {Chi}}, \bibinfo {author} {\bibfnamefont
  {B.}~\bibnamefont {Winn}}, \ and\ \bibinfo {author} {\bibfnamefont
  {P.}~\bibnamefont {Dai}},\ }\href {\doibase 10.1103/PhysRevB.98.060502}
  {\bibfield  {journal} {\bibinfo  {journal} {Phys. Rev. B}\ }\textbf {\bibinfo
  {volume} {98}},\ \bibinfo {pages} {060502} (\bibinfo {year}
  {2018})}\BibitemShut {NoStop}%
\bibitem [{\citenamefont {Kim}\ \emph {et~al.}(2013)\citenamefont {Kim},
  \citenamefont {Tucker}, \citenamefont {Pratt}, \citenamefont {Ran},
  \citenamefont {Thaler}, \citenamefont {Christianson}, \citenamefont {Marty},
  \citenamefont {Calder}, \citenamefont {Podlesnyak}, \citenamefont {Bud'ko},
  \citenamefont {Canfield}, \citenamefont {Kreyssig}, \citenamefont {Goldman},\
  and\ \citenamefont {McQueeney}}]{PhysRevLett.110.177002}%
  \BibitemOpen
  \bibfield  {author} {\bibinfo {author} {\bibfnamefont {M.~G.}\ \bibnamefont
  {Kim}}, \bibinfo {author} {\bibfnamefont {G.~S.}\ \bibnamefont {Tucker}},
  \bibinfo {author} {\bibfnamefont {D.~K.}\ \bibnamefont {Pratt}}, \bibinfo
  {author} {\bibfnamefont {S.}~\bibnamefont {Ran}}, \bibinfo {author}
  {\bibfnamefont {A.}~\bibnamefont {Thaler}}, \bibinfo {author} {\bibfnamefont
  {A.~D.}\ \bibnamefont {Christianson}}, \bibinfo {author} {\bibfnamefont
  {K.}~\bibnamefont {Marty}}, \bibinfo {author} {\bibfnamefont
  {S.}~\bibnamefont {Calder}}, \bibinfo {author} {\bibfnamefont
  {A.}~\bibnamefont {Podlesnyak}}, \bibinfo {author} {\bibfnamefont {S.~L.}\
  \bibnamefont {Bud'ko}}, \bibinfo {author} {\bibfnamefont {P.~C.}\
  \bibnamefont {Canfield}}, \bibinfo {author} {\bibfnamefont {A.}~\bibnamefont
  {Kreyssig}}, \bibinfo {author} {\bibfnamefont {A.~I.}\ \bibnamefont
  {Goldman}}, \ and\ \bibinfo {author} {\bibfnamefont {R.~J.}\ \bibnamefont
  {McQueeney}},\ }\href {\doibase 10.1103/PhysRevLett.110.177002} {\bibfield
  {journal} {\bibinfo  {journal} {Phys. Rev. Lett.}\ }\textbf {\bibinfo
  {volume} {110}},\ \bibinfo {pages} {177002} (\bibinfo {year}
  {2013})}\BibitemShut {NoStop}%
\bibitem [{\citenamefont {Xiao}\ \emph {et~al.}(2009)\citenamefont {Xiao},
  \citenamefont {Su}, \citenamefont {Mittal}, \citenamefont {Chatterji},
  \citenamefont {Hansen}, \citenamefont {Kumar}, \citenamefont {Matsuishi},
  \citenamefont {Hosono},\ and\ \citenamefont {Brueckel}}]{PhysRevB.79.060504}%
  \BibitemOpen
  \bibfield  {author} {\bibinfo {author} {\bibfnamefont {Y.}~\bibnamefont
  {Xiao}}, \bibinfo {author} {\bibfnamefont {Y.}~\bibnamefont {Su}}, \bibinfo
  {author} {\bibfnamefont {R.}~\bibnamefont {Mittal}}, \bibinfo {author}
  {\bibfnamefont {T.}~\bibnamefont {Chatterji}}, \bibinfo {author}
  {\bibfnamefont {T.}~\bibnamefont {Hansen}}, \bibinfo {author} {\bibfnamefont
  {C.~M.~N.}\ \bibnamefont {Kumar}}, \bibinfo {author} {\bibfnamefont
  {S.}~\bibnamefont {Matsuishi}}, \bibinfo {author} {\bibfnamefont
  {H.}~\bibnamefont {Hosono}}, \ and\ \bibinfo {author} {\bibfnamefont
  {T.}~\bibnamefont {Brueckel}},\ }\href {\doibase 10.1103/PhysRevB.79.060504}
  {\bibfield  {journal} {\bibinfo  {journal} {Phys. Rev. B}\ }\textbf {\bibinfo
  {volume} {79}},\ \bibinfo {pages} {060504} (\bibinfo {year}
  {2009})}\BibitemShut {NoStop}%
\bibitem [{\citenamefont {Price}\ \emph {et~al.}(2013)\citenamefont {Price},
  \citenamefont {Su}, \citenamefont {Xiao}, \citenamefont {T.~Adroja},
  \citenamefont {Guidi}, \citenamefont {Mittal}, \citenamefont {Nandi},
  \citenamefont {Matsuishi}, \citenamefont {Hosono},\ and\ \citenamefont
  {Br\"{u}ckel}}]{Price2013}%
  \BibitemOpen
  \bibfield  {author} {\bibinfo {author} {\bibfnamefont {S.}~\bibnamefont
  {Price}}, \bibinfo {author} {\bibfnamefont {Y.}~\bibnamefont {Su}}, \bibinfo
  {author} {\bibfnamefont {Y.}~\bibnamefont {Xiao}}, \bibinfo {author}
  {\bibfnamefont {D.}~\bibnamefont {T.~Adroja}}, \bibinfo {author}
  {\bibfnamefont {T.}~\bibnamefont {Guidi}}, \bibinfo {author} {\bibfnamefont
  {R.}~\bibnamefont {Mittal}}, \bibinfo {author} {\bibfnamefont
  {S.}~\bibnamefont {Nandi}}, \bibinfo {author} {\bibfnamefont
  {S.}~\bibnamefont {Matsuishi}}, \bibinfo {author} {\bibfnamefont
  {H.}~\bibnamefont {Hosono}}, \ and\ \bibinfo {author} {\bibfnamefont
  {T.}~\bibnamefont {Br\"{u}ckel}},\ }\href {\doibase 10.7566/JPSJ.82.104716}
  {\bibfield  {journal} {\bibinfo  {journal} {Journal of the Physical Society
  of Japan}\ }\textbf {\bibinfo {volume} {82}},\ \bibinfo {pages} {104716}
  (\bibinfo {year} {2013})}\BibitemShut {NoStop}%
\bibitem [{\citenamefont {Ma}\ \emph {et~al.}(2022)\citenamefont {Ma},
  \citenamefont {Ruan}, \citenamefont {Zhou}, \citenamefont {Gu}, \citenamefont
  {Yang}, \citenamefont {Sun},\ and\ \citenamefont {Ren}}]{MA2022126562}%
  \BibitemOpen
  \bibfield  {author} {\bibinfo {author} {\bibfnamefont {M.-W.}\ \bibnamefont
  {Ma}}, \bibinfo {author} {\bibfnamefont {B.}~\bibnamefont {Ruan}}, \bibinfo
  {author} {\bibfnamefont {M.}~\bibnamefont {Zhou}}, \bibinfo {author}
  {\bibfnamefont {Y.}~\bibnamefont {Gu}}, \bibinfo {author} {\bibfnamefont
  {Q.}~\bibnamefont {Yang}}, \bibinfo {author} {\bibfnamefont {J.}~\bibnamefont
  {Sun}}, \ and\ \bibinfo {author} {\bibfnamefont {Z.-A.}\ \bibnamefont
  {Ren}},\ }\href {\doibase https://doi.org/10.1016/j.jcrysgro.2022.126562}
  {\bibfield  {journal} {\bibinfo  {journal} {Journal of Crystal Growth}\
  }\textbf {\bibinfo {volume} {585}},\ \bibinfo {pages} {126562} (\bibinfo
  {year} {2022})}\BibitemShut {NoStop}%
\bibitem [{\citenamefont {Sidis}\ \emph {et~al.}(2004)\citenamefont {Sidis},
  \citenamefont {Pailh\`es}, \citenamefont {Keimer}, \citenamefont {Bourges},
  \citenamefont {Ulrich},\ and\ \citenamefont {Regnault}}]{SidisY}%
  \BibitemOpen
  \bibfield  {author} {\bibinfo {author} {\bibfnamefont {Y.}~\bibnamefont
  {Sidis}}, \bibinfo {author} {\bibfnamefont {S.}~\bibnamefont {Pailh\`es}},
  \bibinfo {author} {\bibfnamefont {B.}~\bibnamefont {Keimer}}, \bibinfo
  {author} {\bibfnamefont {P.}~\bibnamefont {Bourges}}, \bibinfo {author}
  {\bibfnamefont {C.}~\bibnamefont {Ulrich}}, \ and\ \bibinfo {author}
  {\bibfnamefont {L.~P.}\ \bibnamefont {Regnault}},\ }\href {\doibase
  https://doi.org/10.1002/pssb.200304498} {\bibfield  {journal} {\bibinfo
  {journal} {physica status solidi (b)}\ }\textbf {\bibinfo {volume} {241}},\
  \bibinfo {pages} {1204} (\bibinfo {year} {2004})}\BibitemShut {NoStop}%
\bibitem [{\citenamefont {Inosov}\ \emph {et~al.}(2010)\citenamefont {Inosov},
  \citenamefont {Park}, \citenamefont {Bourges}, \citenamefont {Sun},
  \citenamefont {Sidis}, \citenamefont {Schneidewind}, \citenamefont {Hradil},
  \citenamefont {Haug}, \citenamefont {Lin}, \citenamefont {Keimer},\ and\
  \citenamefont {Hinkov}}]{InosovNaturePhys2010}%
  \BibitemOpen
  \bibfield  {author} {\bibinfo {author} {\bibfnamefont {D.~S.}\ \bibnamefont
  {Inosov}}, \bibinfo {author} {\bibfnamefont {J.~T.}\ \bibnamefont {Park}},
  \bibinfo {author} {\bibfnamefont {P.}~\bibnamefont {Bourges}}, \bibinfo
  {author} {\bibfnamefont {D.~L.}\ \bibnamefont {Sun}}, \bibinfo {author}
  {\bibfnamefont {Y.}~\bibnamefont {Sidis}}, \bibinfo {author} {\bibfnamefont
  {A.}~\bibnamefont {Schneidewind}}, \bibinfo {author} {\bibfnamefont
  {K.}~\bibnamefont {Hradil}}, \bibinfo {author} {\bibfnamefont
  {D.}~\bibnamefont {Haug}}, \bibinfo {author} {\bibfnamefont {C.~T.}\
  \bibnamefont {Lin}}, \bibinfo {author} {\bibfnamefont {B.}~\bibnamefont
  {Keimer}}, \ and\ \bibinfo {author} {\bibfnamefont {V.}~\bibnamefont
  {Hinkov}},\ }\href {\doibase https://doi.org/10.1038/nphys1483} {\bibfield
  {journal} {\bibinfo  {journal} {Nat. Phys.}\ }\textbf {\bibinfo {volume}
  {6}},\ \bibinfo {pages} {178} (\bibinfo {year} {2010})}\BibitemShut {NoStop}%
\bibitem [{\citenamefont {Xie}\ \emph {et~al.}(2018)\citenamefont {Xie},
  \citenamefont {Gong}, \citenamefont {Ghosh}, \citenamefont {Ghosh},
  \citenamefont {Soda}, \citenamefont {Masuda}, \citenamefont {Itoh},
  \citenamefont {Bourdarot}, \citenamefont {Regnault}, \citenamefont
  {Danilkin}, \citenamefont {Li},\ and\ \citenamefont
  {Luo}}]{PhysRevLett.120.137001}%
  \BibitemOpen
  \bibfield  {author} {\bibinfo {author} {\bibfnamefont {T.}~\bibnamefont
  {Xie}}, \bibinfo {author} {\bibfnamefont {D.}~\bibnamefont {Gong}}, \bibinfo
  {author} {\bibfnamefont {H.}~\bibnamefont {Ghosh}}, \bibinfo {author}
  {\bibfnamefont {A.}~\bibnamefont {Ghosh}}, \bibinfo {author} {\bibfnamefont
  {M.}~\bibnamefont {Soda}}, \bibinfo {author} {\bibfnamefont {T.}~\bibnamefont
  {Masuda}}, \bibinfo {author} {\bibfnamefont {S.}~\bibnamefont {Itoh}},
  \bibinfo {author} {\bibfnamefont {F.}~\bibnamefont {Bourdarot}}, \bibinfo
  {author} {\bibfnamefont {L.-P.}\ \bibnamefont {Regnault}}, \bibinfo {author}
  {\bibfnamefont {S.}~\bibnamefont {Danilkin}}, \bibinfo {author}
  {\bibfnamefont {S.}~\bibnamefont {Li}}, \ and\ \bibinfo {author}
  {\bibfnamefont {H.}~\bibnamefont {Luo}},\ }\href {\doibase
  10.1103/PhysRevLett.120.137001} {\bibfield  {journal} {\bibinfo  {journal}
  {Phys. Rev. Lett.}\ }\textbf {\bibinfo {volume} {120}},\ \bibinfo {pages}
  {137001} (\bibinfo {year} {2018})}\BibitemShut {NoStop}%
\bibitem [{\citenamefont {Hong}\ \emph {et~al.}(2020)\citenamefont {Hong},
  \citenamefont {Song}, \citenamefont {Liu}, \citenamefont {Li}, \citenamefont
  {Zeng}, \citenamefont {Li}, \citenamefont {Wu}, \citenamefont {Sui},
  \citenamefont {Xie}, \citenamefont {Danilkin}, \citenamefont {Ghosh},
  \citenamefont {Ghosh}, \citenamefont {Hu}, \citenamefont {Zhao},
  \citenamefont {Zhou}, \citenamefont {Qiu}, \citenamefont {Li},\ and\
  \citenamefont {Luo}}]{PhysRevLett.125.117002}%
  \BibitemOpen
  \bibfield  {author} {\bibinfo {author} {\bibfnamefont {W.}~\bibnamefont
  {Hong}}, \bibinfo {author} {\bibfnamefont {L.}~\bibnamefont {Song}}, \bibinfo
  {author} {\bibfnamefont {B.}~\bibnamefont {Liu}}, \bibinfo {author}
  {\bibfnamefont {Z.}~\bibnamefont {Li}}, \bibinfo {author} {\bibfnamefont
  {Z.}~\bibnamefont {Zeng}}, \bibinfo {author} {\bibfnamefont {Y.}~\bibnamefont
  {Li}}, \bibinfo {author} {\bibfnamefont {D.}~\bibnamefont {Wu}}, \bibinfo
  {author} {\bibfnamefont {Q.}~\bibnamefont {Sui}}, \bibinfo {author}
  {\bibfnamefont {T.}~\bibnamefont {Xie}}, \bibinfo {author} {\bibfnamefont
  {S.}~\bibnamefont {Danilkin}}, \bibinfo {author} {\bibfnamefont
  {H.}~\bibnamefont {Ghosh}}, \bibinfo {author} {\bibfnamefont
  {A.}~\bibnamefont {Ghosh}}, \bibinfo {author} {\bibfnamefont
  {J.}~\bibnamefont {Hu}}, \bibinfo {author} {\bibfnamefont {L.}~\bibnamefont
  {Zhao}}, \bibinfo {author} {\bibfnamefont {X.}~\bibnamefont {Zhou}}, \bibinfo
  {author} {\bibfnamefont {X.}~\bibnamefont {Qiu}}, \bibinfo {author}
  {\bibfnamefont {S.}~\bibnamefont {Li}}, \ and\ \bibinfo {author}
  {\bibfnamefont {H.}~\bibnamefont {Luo}},\ }\href {\doibase
  10.1103/PhysRevLett.125.117002} {\bibfield  {journal} {\bibinfo  {journal}
  {Phys. Rev. Lett.}\ }\textbf {\bibinfo {volume} {125}},\ \bibinfo {pages}
  {117002} (\bibinfo {year} {2020})}\BibitemShut {NoStop}%
\bibitem [{\citenamefont {Zhang}\ \emph {et~al.}(2010)\citenamefont {Zhang},
  \citenamefont {Sknepnek},\ and\ \citenamefont
  {Schmalian}}]{PhysRevB.82.134527}%
  \BibitemOpen
  \bibfield  {author} {\bibinfo {author} {\bibfnamefont {J.}~\bibnamefont
  {Zhang}}, \bibinfo {author} {\bibfnamefont {R.}~\bibnamefont {Sknepnek}}, \
  and\ \bibinfo {author} {\bibfnamefont {J.}~\bibnamefont {Schmalian}},\ }\href
  {\doibase 10.1103/PhysRevB.82.134527} {\bibfield  {journal} {\bibinfo
  {journal} {Phys. Rev. B}\ }\textbf {\bibinfo {volume} {82}},\ \bibinfo
  {pages} {134527} (\bibinfo {year} {2010})}\BibitemShut {NoStop}%
\bibitem [{\citenamefont {Maier}\ \emph {et~al.}(2012)\citenamefont {Maier},
  \citenamefont {Hirschfeld},\ and\ \citenamefont
  {Scalapino}}]{PhysRevB.86.094514}%
  \BibitemOpen
  \bibfield  {author} {\bibinfo {author} {\bibfnamefont {T.~A.}\ \bibnamefont
  {Maier}}, \bibinfo {author} {\bibfnamefont {P.~J.}\ \bibnamefont
  {Hirschfeld}}, \ and\ \bibinfo {author} {\bibfnamefont {D.~J.}\ \bibnamefont
  {Scalapino}},\ }\href {\doibase 10.1103/PhysRevB.86.094514} {\bibfield
  {journal} {\bibinfo  {journal} {Phys. Rev. B}\ }\textbf {\bibinfo {volume}
  {86}},\ \bibinfo {pages} {094514} (\bibinfo {year} {2012})}\BibitemShut
  {NoStop}%
\bibitem [{\citenamefont {Das}\ and\ \citenamefont
  {Balatsky}(2011)}]{PhysRevLett.106.157004}%
  \BibitemOpen
  \bibfield  {author} {\bibinfo {author} {\bibfnamefont {T.}~\bibnamefont
  {Das}}\ and\ \bibinfo {author} {\bibfnamefont {A.~V.}\ \bibnamefont
  {Balatsky}},\ }\href {\doibase 10.1103/PhysRevLett.106.157004} {\bibfield
  {journal} {\bibinfo  {journal} {Phys. Rev. Lett.}\ }\textbf {\bibinfo
  {volume} {106}},\ \bibinfo {pages} {157004} (\bibinfo {year}
  {2011})}\BibitemShut {NoStop}%
\bibitem [{\citenamefont {Zhang}\ \emph {et~al.}(2015)\citenamefont {Zhang},
  \citenamefont {Fernandes}, \citenamefont {Lamsal}, \citenamefont {Yan},
  \citenamefont {Chi}, \citenamefont {Tucker}, \citenamefont {Pratt},
  \citenamefont {Lynn}, \citenamefont {McCallum}, \citenamefont {Canfield},
  \citenamefont {Lograsso}, \citenamefont {Goldman}, \citenamefont {Vaknin},\
  and\ \citenamefont {McQueeney}}]{PhysRevLett.114.057001}%
  \BibitemOpen
  \bibfield  {author} {\bibinfo {author} {\bibfnamefont {Q.}~\bibnamefont
  {Zhang}}, \bibinfo {author} {\bibfnamefont {R.~M.}\ \bibnamefont
  {Fernandes}}, \bibinfo {author} {\bibfnamefont {J.}~\bibnamefont {Lamsal}},
  \bibinfo {author} {\bibfnamefont {J.}~\bibnamefont {Yan}}, \bibinfo {author}
  {\bibfnamefont {S.}~\bibnamefont {Chi}}, \bibinfo {author} {\bibfnamefont
  {G.~S.}\ \bibnamefont {Tucker}}, \bibinfo {author} {\bibfnamefont {D.~K.}\
  \bibnamefont {Pratt}}, \bibinfo {author} {\bibfnamefont {J.~W.}\ \bibnamefont
  {Lynn}}, \bibinfo {author} {\bibfnamefont {R.~W.}\ \bibnamefont {McCallum}},
  \bibinfo {author} {\bibfnamefont {P.~C.}\ \bibnamefont {Canfield}}, \bibinfo
  {author} {\bibfnamefont {T.~A.}\ \bibnamefont {Lograsso}}, \bibinfo {author}
  {\bibfnamefont {A.~I.}\ \bibnamefont {Goldman}}, \bibinfo {author}
  {\bibfnamefont {D.}~\bibnamefont {Vaknin}}, \ and\ \bibinfo {author}
  {\bibfnamefont {R.~J.}\ \bibnamefont {McQueeney}},\ }\href {\doibase
  10.1103/PhysRevLett.114.057001} {\bibfield  {journal} {\bibinfo  {journal}
  {Phys. Rev. Lett.}\ }\textbf {\bibinfo {volume} {114}},\ \bibinfo {pages}
  {057001} (\bibinfo {year} {2015})}\BibitemShut {NoStop}%
\bibitem [{\citenamefont {Zhang}\ \emph {et~al.}(2016)\citenamefont {Zhang},
  \citenamefont {Park}, \citenamefont {Lu}, \citenamefont {Wei}, \citenamefont
  {Ma}, \citenamefont {Hao}, \citenamefont {Dai}, \citenamefont {Meng},
  \citenamefont {Yang}, \citenamefont {Luo},\ and\ \citenamefont
  {Li}}]{PhysRevLett.117.227003}%
  \BibitemOpen
  \bibfield  {author} {\bibinfo {author} {\bibfnamefont {W.}~\bibnamefont
  {Zhang}}, \bibinfo {author} {\bibfnamefont {J.~T.}\ \bibnamefont {Park}},
  \bibinfo {author} {\bibfnamefont {X.}~\bibnamefont {Lu}}, \bibinfo {author}
  {\bibfnamefont {Y.}~\bibnamefont {Wei}}, \bibinfo {author} {\bibfnamefont
  {X.}~\bibnamefont {Ma}}, \bibinfo {author} {\bibfnamefont {L.}~\bibnamefont
  {Hao}}, \bibinfo {author} {\bibfnamefont {P.}~\bibnamefont {Dai}}, \bibinfo
  {author} {\bibfnamefont {Z.~Y.}\ \bibnamefont {Meng}}, \bibinfo {author}
  {\bibfnamefont {Y.-f.}\ \bibnamefont {Yang}}, \bibinfo {author}
  {\bibfnamefont {H.}~\bibnamefont {Luo}}, \ and\ \bibinfo {author}
  {\bibfnamefont {S.}~\bibnamefont {Li}},\ }\href {\doibase
  10.1103/PhysRevLett.117.227003} {\bibfield  {journal} {\bibinfo  {journal}
  {Phys. Rev. Lett.}\ }\textbf {\bibinfo {volume} {117}},\ \bibinfo {pages}
  {227003} (\bibinfo {year} {2016})}\BibitemShut {NoStop}%
\bibitem [{\citenamefont {Wang}\ \emph {et~al.}(2013)\citenamefont {Wang},
  \citenamefont {Zhang}, \citenamefont {Lu}, \citenamefont {Tan}, \citenamefont
  {Luo}, \citenamefont {Song}, \citenamefont {Wang}, \citenamefont {Zhang},
  \citenamefont {Goremychkin}, \citenamefont {Perring}, \citenamefont {Maier},
  \citenamefont {Yin}, \citenamefont {Haule}, \citenamefont {Kotliar},\ and\
  \citenamefont {Dai}}]{WangMNC}%
  \BibitemOpen
  \bibfield  {author} {\bibinfo {author} {\bibfnamefont {M.}~\bibnamefont
  {Wang}}, \bibinfo {author} {\bibfnamefont {C.}~\bibnamefont {Zhang}},
  \bibinfo {author} {\bibfnamefont {X.}~\bibnamefont {Lu}}, \bibinfo {author}
  {\bibfnamefont {G.}~\bibnamefont {Tan}}, \bibinfo {author} {\bibfnamefont
  {H.}~\bibnamefont {Luo}}, \bibinfo {author} {\bibfnamefont {Y.}~\bibnamefont
  {Song}}, \bibinfo {author} {\bibfnamefont {M.}~\bibnamefont {Wang}}, \bibinfo
  {author} {\bibfnamefont {X.}~\bibnamefont {Zhang}}, \bibinfo {author}
  {\bibfnamefont {E.}~\bibnamefont {Goremychkin}}, \bibinfo {author}
  {\bibfnamefont {T.}~\bibnamefont {Perring}}, \bibinfo {author} {\bibfnamefont
  {T.}~\bibnamefont {Maier}}, \bibinfo {author} {\bibfnamefont
  {Z.}~\bibnamefont {Yin}}, \bibinfo {author} {\bibfnamefont {K.}~\bibnamefont
  {Haule}}, \bibinfo {author} {\bibfnamefont {G.}~\bibnamefont {Kotliar}}, \
  and\ \bibinfo {author} {\bibfnamefont {P.}~\bibnamefont {Dai}},\ }\href
  {\doibase https://doi.org/10.1038/ncomms3874} {\bibfield  {journal} {\bibinfo
   {journal} {Nat. Commun.}\ }\textbf {\bibinfo {volume} {4}},\ \bibinfo
  {pages} {2874} (\bibinfo {year} {2013})}\BibitemShut {NoStop}%
\end{thebibliography}

\end{document}